\documentclass[12pt,notitlepage,a4paper]{article}
\usepackage{color,graphicx}
\usepackage{cite}
\usepackage{amssymb}
\usepackage{delarray,amsmath}
\usepackage[latin1]{inputenc}
\usepackage[american]{babel}
\usepackage[table]{xcolor}
\usepackage{cite}
\usepackage{tikz-cd}
\usepackage{hyperref}

\newcommand{\be}{\begin{equation}}
\newcommand{\ee}{\end{equation}}
\newcommand{\beq}{\begin{equation}}
\newcommand{\eeq}{\end{equation}}
\newcommand{\beqa}{\begin{eqnarray}}
\newcommand{\eeqa}{\end{eqnarray}}

\newcommand{\bear}{\begin{eqnarray}}
\newcommand{\eear}{\end{eqnarray}}

\numberwithin{equation}{section}

\newfont{\namefont}{cmr10}
\newfont{\addfont}{cmti7 scaled 1440}
\newfont{\boldmathfont}{cmbx10}
\newfont{\headfontb}{cmbx10 scaled 1728}

\begin{document}
\baselineskip=15.5pt
\pagestyle{plain}
\setcounter{page}{1}

\begin{center}
\vspace{0.1in}

\renewcommand{\thefootnote}{\fnsymbol{footnote}}

\begin{center}
\Large \bf    Gravitational wave driving of a \\ gapped holographic system 
\end{center}
\vskip 0.1truein
\begin{center}
\bf{Anxo Biasi${}^1$\footnote{anxo.biasi@gmail.com}, 
Javier Mas${}^1$\footnote{javier.mas@usc.es} and 
Alexandre Serantes${}^3$\footnote{alexandre.serantes@icts.res.in}}

\end{center}
\vspace{0.5mm}

\begin{center}\it{
${}^1$Departamento de  F\'\i sica de Part\'\i  culas \\
Universidade de Santiago de Compostela \\
and \\
Instituto Galego de F\'\i sica de Altas Enerx\'\i as (IGFAE)\\
E-15782 Santiago de Compostela, Spain}
\end{center}

\begin{center}\it{
${}^3$}International Centre for Theoretical Sciences-TIFR, \\
Survey No. 151, Shivakote, Hesaraghatta Hobli, \\
Bengaluru North, India 560 089
\end{center}

\setcounter{footnote}{0}
\renewcommand{\thefootnote}{\arabic{footnote}}

\vspace{0.4in}

\begin{abstract}
\noindent
This work addresses the response of a holographic conformal field theory to a homogeneous gravitational periodic driving. The dual geometry is the 
AdS-soliton, which models a strongly coupled quantum system in a gapped phase, on a compact domain. The response is a time-periodic geometry up to a driving
amplitude threshold which decreases with the driving frequency. Beyond that, collapse to a black hole occurs, signaling decoherence and thermalization in the dual theory. 
At some frequencies, we also find a resonant coupling to the gravitational normal modes of the AdS-soliton, yielding a nonlinearly bound state.
 We also speculate on the possible uses of quantum strongly coupled systems to build resonant gravitational wave detectors.

\smallskip
\end{abstract}
\end{center}

\newpage

\section{Introduction}

A periodic drive is one of the simplest, yet most fascinating, ways of taking a many-body quantum system out of equilibrium. The discrete time-translational invariance retained by the drive has crucial consequences for the description of the unitary evolution of the system; in particular, at stroboscopic times, the dynamics is controlled by an emergent, time-independent hermitian operator: the Floquet Hamiltonian. This observation opens the possibility of driving otherwise autonomous systems to manufacture Floquet hamiltonians of physical relevance. The topic, which goes under the name of Floquet engineering, has been under intense scrutiny  in recent years (see \cite{eckardt2017colloquium}\cite{oka2018floquet} \cite{bukov2015universal} \cite{weinberg2017adiabatic} for reviews). At the same time, fundamental questions regarding the late-time behavior of periodically driven, many-body quantum systems have also been thoroughly studied \cite{d2014long}\cite{lazarides2014equilibrium}\cite{ponte2015periodically}. 

At the QFT level, less is known in comparison, although remarkable results have been obtained for scalar field theories with O($N$) symmetry at large N \cite{PhysRevB.93.174305}\cite{weidinger2017floquet}. In the large N limit, Holography is firmly established as a first-principles framework to deal with real-time physical problems in strongly coupled CFTs. Therefore, it provides an interesting starting point to increase our knowledge about periodically driven QFTs, a possibility that has not gone unnoticed \cite{Li:2013fhw}\cite{auzzi2013periodically}\cite{rangamani2015driven}\cite{Hashimoto:2016ize}\cite{Kinoshita:2017uch}\cite{Ishii:2018ucz}\cite{haack2018probing}.  

In \cite{biasi2018floquet} the problem of periodically driven, finite-size systems in the holographic context was explored. The construction involved a CFT placed on a two-sphere, where the drive was implemented by turning on a spatially homogeneous and time-harmonic coupling to a marginal scalar operator. The main reason for considering such a setup was to address the effect of the periodic excitation on a pure state, searching in particular for nontrivial late-time dynamics. Several phenomena were found, such as the existence of exactly time-periodic geometries, dynamical phase transitions, or hysteresis loops. 

The present paper aims at furthering  this subject. One of the important questions that will guide our work is elucidating whether the physics uncovered in \cite{biasi2018floquet} is specific to that particular context or, on the other hand, displays some degree of universality. To this end, we will consider another example of a finite-sized,  two-dimensional system: a holographic CFT placed on a two-torus. When imposing supersymmetry-breaking boundary conditions along each cycle of the torus, the ground state of these theories is known to be dual to the AdS-soliton geometry \cite{horowitz1998ads}\cite{Witten:1998zw}, which is nothing but the double analytical continuation of a planar AdS-Schwarzschild black hole. 
%
%

Our playground will be the four-dimensional instance of this geometry, which will be subjected to a time-periodic shear deformation of its boundary metric. The reason for considering this particular setup is two-fold. On the one hand, it allows us to remain confined to a universal subsector of the AdS/CFT correspondence involving the dynamics of the metric/energy-momentum tensor alone; in this way, the results we will obtain pertain to any holographic CFT that respects our boundary conditions. 
On the other, this setup can be viewed as a toy-model of the response of a table-top quantum system to the passage of a gravitational wave \cite{sabin2014phonon} \cite{Schutzhold:2018wfu}\cite{Robbins:2018thb}. The toroidal topology accounts for the periodic boundary conditions in a rectangular-shaped, strongly coupled quantum system, pierced by a gravitational wave propagating in the perpendicular direction. Related work, studying the evolution of the AdS-soliton geometry after quantum quenches, can be found in \cite{Craps:2015upq}\cite{Myers:2017sxr}. 

This paper is organized as follows. In section \ref{sec1} we describe our model. Then, in section \ref{sec2}, we demonstrate the existence of exactly time-periodic geometries by construction. We chart out their phase diagram and discuss their linear and nonlinear stability properties in detail. Section \ref{sec3} is devoted to the study of modulated driving protocols. These are processes in which the parameters determining the driving become themselves nontrivial functions of time. Thus, they allow for a dynamical interpolation between the undriven ground state of our system and a time-periodic geometry. First, we discuss the system response to such boundary conditions in the quasi-static limit, studying in detail under which circumstances the response is adiabatic. Then, we illustrate what happens away from the quasi-static limit, showing that specific kinds of dynamical phase transitions might take place. The manuscript closes down in section \ref{sec4} with a summary of our main results, their possible implications for gravitational wave detection, and the suggestion of future avenues worth exploring.   

\section{The setup}
\label{sec1}

In this paper, we consider pure Einstein gravity in four-dimensions with a negative cosmological constant and study the behavior of an AdS-soliton background under a time-dependent strained boundary metric. The metric ansatz,  
\begin{equation}
ds^2 = \frac{L^2}{z^2} \left( - f(t,z) e^{-2\delta(t,z)} dt^2 + \frac{dz^2}{\left( 1-\displaystyle\frac{z^3}{z_0^3}\right) f(t,z)} + \left(1-\frac{z^3}{z_0^3}\right) e^{ b(t,z)} dx^2 + e^{-b(t,z)}dy^2 \right)\, ,
\label{metric}
\end{equation}
generalizes the AdS-soliton geometry,  which is recovered for $b(t,z) = \delta(t,z) = 0$, $f(t,z) = 1$.  A nonzero $b(t,z)$ introduces an anisotropy in the boundary geometry; for small amplitudes,  it acts like the $h_+$ strain of a gravitational wave propagating along the direction perpendicular to the dual quantum system. 

$L$ is the AdS radius and we set it to $1$ by using it as our length stick. The holographic coordinate $z$  ranges from $z=0$ at the asymptotic boundary to $z=z_0$ at the tip of the cigar, where the proper length along the $x$ direction vanishes and spacetime ends. Regularity at this tip requires $x$ to be a compact coordinate with periodicity $L_x = 4\pi z_0/3$. It is a very satisfactory feature of this solution to see the deep bulk bound $z  \leq z_0$ reflecting itself as an IR cutoff in the boundary ($x<L_x$), much in the spirit of the holographic UV/IR  correspondence \cite{Peet:1998wn}. It would be nice to have similar arguments at work for the other coordinate $y$, enforcing on it a periodicity $L_y$ by singularity avoidance at a deeper bulk scale $z_1\geq z_0$. Unfortunately, no such solution is known.\footnote{See however \cite{Myers:1999psa} for a discussion on a singular geometry that has some of the desired ingredients. The  first order phase transition upon exchange of periodicities have been discussed in several places, see for example  \cite{Page:2002qc}\cite{Belin:2016yll}\cite{Bueno:2016rma}.
}
In our setup, the coordinate $y$ can be compactified for free, as long as its period $L_y$  is longer than $L_x$. Thus, we will envisage the geometry (\ref{metric}) as being dual to a strongly coupled CFT placed on a rectangular region with sides  $L_y> L_x$ and periodic/antiperiodic boundary conditions for bosonic/fermionic fields. It is on this setup that our investigations aim at unveiling detectable effects of the gravitational driving parametrized by the strain field $b(t,z)$. 
 
The equations of motion and boundary conditions to be imposed at the tip can be found in Appendix \ref{app_A}. Here, we will focus solely on reviewing some relevant aspects of the model.
 First, close to the asymptotic boundary, $z = 0$, the ultraviolet expansion of the metric reads
\beqa
\delta(t,z) &=& \frac{1}{8}  \dot b_0(t)^2  z^2 + {\cal O}(z^4), \nonumber \\
f(t,z) &=& 1 - \frac{1}{4}   \dot b_0(t)^2 z^2  + f_3(t) z^3 + {\cal O}(z^4), \nonumber\\
b(t,z) &=& b_0(t) -\frac{1}{2} \ddot b_0(t) z^2 + b_3(t) z^3 + {\cal O}(z^4), 
\eeqa
where we have selected the boundary proper time gauge $\delta(t,0) = 0$. We observe that there are three boundary functions left free in the Frobenius expansion: $b_0(t), f_3(t)$ and $b_3(t)$. They are linked by the momentum constraint equation \eqref{momcons} 
 \be
 \dot f_3(t) = \frac{3}{2} b_3(t)\dot b_0(t) -\frac{3}{4 z_0^3} \dot b_0(t). \label{WI}
 \ee
The holographic dictionary relates these quantities to field theory data as follows
\beqa
h_{ab}  &=&  {\rm diag}\left[  -1, e^{b_0(t)}, e^{-b_0(t)} \right] \label{boundarydata}\\
\langle T_{ab}\rangle  &=&  \frac{1}{16\pi G_4} {\rm diag} \left[ - z_0^{-3} - 2 f_3(t) ,e^{b_0(t)}(-2 z_0^{-3}-f_3(t) + 3 b_3(t)),  e^{-b_0(t)}(z_0^{-3}-f_3(t) - 3 b_3(t))  \right] \, , \nonumber
\eeqa
where $h_{ab}$ is the induced metric of the boundary theory and $T_{ab}$ its energy-momentum tensor. Under this perspective, equation (\ref{WI}) is nothing but the statement of energy-momentum conservation, $\nabla_a \langle T^{ab}\rangle = 0$,  which follows from the reparametrization invariance of the bulk theory \cite{deHaro:2000vlm}.\footnote{ The covariant derivative $\nabla_a$ is to be understood as defined with respect to the time-dependent boundary metric $h_{ab}$.} 

From $\langle T_{00}\rangle$,  we see that $f_3$ controls the energy density of the dual CFT. Hence, \eqref{WI} states that, as soon as $\dot b_0\neq 0$,   there is energy exchange across the boundary and we are dealing with an open system. In this paper,  we will focus on a harmonic driving of frequency $\omega_b$ measured with respect to the proper time coordinate at the boundary\footnote{The boundary-time gauge choice $\delta(t.z=0) = 0$ is precisely ensuring this fact.}
\begin{equation}
b_0(t) =\rho \cos (\omega_b t). \label{harmonic_driving}
\end{equation}
Looking again at equation \eqref{WI}, we notice that the last term, which only contains $\dot b_0(t)$,  averages to zero over a driving period. The fate of the mixed product $\dot b_0(t) b_3(t)$ is, on the other hand, unpredictable. Since it does not have a definite sign, the system can undergo energy exchanges of both signs. The late-time evolution of the system is linked to this term, whose knowledge demands an integration over the radial domain in order to extract $b_3(t)$ at each instant of time. 

 Let us close this section by pointing out that, despite the fact that $z_0$ will show up in the main body of the paper when referring to the position of the tip of the cigar in the holographic direction, numerical results and plots are given for $z_0=1$ without loss of generality. This is, as usual, not a choice of scale, that one being fixed once and for all by setting $L=1$, or else, by measuring all lengths in units of $L$. It rather comes from the residual  symmetry of the ansatz \eqref{metric} under rescaling $x^A \to \lambda x^A$, where $x^A = (t,z,x,y)$, together with $z_0\to \lambda z_0$. Solutions can be mapped to one another by this scaling, and distinct families are labeled by invariant quotients of the form $t/z_0$, etc. Inequivalent drivings, henceforth, are parametrized by  the dimensionless frequency $\omega = \omega_b z_0$, as follows from writing $\cos (\omega_b t) = \cos(\omega t/z_0)$.

\section{Time-periodic solutions}
\label{sec2}

It is natural to start looking for solutions that inherit the exact periodicity of the harmonic driving \eqref{harmonic_driving}. We will refer to these geometries as \textit{time-periodic solutions} (TPSs). The numerical procedure needed to construct them follows  the methods developed for a scalar field in global AdS \cite{biasi2018floquet}\cite{Maliborski:2013jca}\cite{Maliborski:2016zlh}\cite{Carracedo:2016qrf}.
The strategy starts by first elucidating, through a perturbative analysis, the expected content of Fourier harmonics of a TPS. Inserting the ansatz 
\be
 b(t,z) =\sum_{n = 1}^{\infty} \epsilon^n b_n(t,z)~~,~~P(t,z) = \sum_{n = 1}^{\infty} \epsilon^n P_n(t,z) ~~, ~~ \omega = \omega_0 + \sum_{i=1}^\infty \epsilon^i \omega_i
\ee
into the equations of motion  \eqref{eq_B_P_z}-\eqref{eq_delta_z}, the system can be solved in a power series in $\epsilon$, with boundary conditions $b_n(t,0)= \delta_{n,1}\rho \cos \omega t$ and $ P_n(t,0) = -\delta_{n,1}\omega  \rho \sin \omega t$. The upshot is that, at $n$-th order, $b(t,z)$ contains {\em all} cosine functions $\cos(k\omega t)$ with $k=1,2,...n$. In other words, all multiples of the driving frequency $\omega$ give rise to Fourier modes of the nonlinear solutions branching off a linearized mode $b_1(t,z)$. This is in sharp contrast with the case of a real massless scalar field, and follows from the lack of reflection symmetry $b \to - b$ of our problem. 
Armed with this result, we can infer the general structure of a TPS 
\beq
b(t,z) = \sum_{k=1}^{\infty}\cos(k \omega t) b_{k}(z),~~~~~~P(t,z) = \sum_{k=1}^{\infty}\sin(k \omega t) p_{k}(z),  
\label{eq_TPS}
\eeq
where the asymptotic boundary conditions are $b_{k}(0) =  \rho \delta_{k,1}$ and $p_{k}(0) =  -\omega \rho \delta_{k,1}$. Finally, a version of the ansatz \eqref{eq_TPS} -truncated to $k\leq K \in \mathbb N$- can be solved numerically by a pseudospectral  method in a  spacetime rectangular domain. 

The TPSs with vanishing driving amplitude, $\rho = 0$, are of particular interest. At first order in $\epsilon$,  nontrivial solutions for $b_{1}(z)$ only exist for a discrete set of frequencies $\omega$, yielding the normal mode spectrum $\Omega_n \approx  2.149,4.790,7.116, 9.389...$ \cite{Craps:2015upq}. For finite amplitudes, these solutions can be uplifted to {\em nonlinear gravitational normal modes}. These sourceless oscillating solutions, termed oscillons (boson stars) in the context of a real (complex) scalar field in AdS, appear as important landmarks in the phase diagram of solutions.

From the dual field theory perspective, it is natural to conjecture that TPSs correspond to {\em Floquet condensates} \cite{Kinoshita:2017uch}. This concept was introduced in the context of the periodically driven two-site Bose-Hubbard model \cite{2016JMOp...63.1768H}, where they are defined as many-body states that share the coherence properties of a mesoscopically occupied, single-particle state of an effective (Floquet) Hamiltonian. Pursuing the analogy, we will envisage our exactly periodic geometries as providing the holographic dual of such Floquet condensates, since these horizonless time-periodic solutions correspond to pure excited states in dual CFT with a macroscopic energy density that are perfectly synchronized with the drive. 
%
%

After constructing the fully nonlinear TPSs, we can employ them to explore the phase space of our system. To see where to expect a transition to other type of solutions we begin by inspecting the linear stability of the TPSs.  Consider the following ansatz for the  linearized fluctuations
\be
\psi(t,z) = \psi_p(t,z) + \epsilon\,\tilde\psi(t,z) \label{psipert}
\ee

where $ \psi = \{b,P,f,\delta \}$ 
and $\psi_p(t,z)$  stands for an exact TPS solution en each case. Inserting 
\eqref{psipert} in \eqref{eq_B_P_z}-\eqref{eq_delta_z} and expanding in $\epsilon$, we obtain the equations of motion for the linear perturbations. The fluctuations are chosen to vanish at the boundary,  
$\tilde\psi(t,0) = 0$, in order not to modify the asymptotic structure of the spacetime, nor the driving. Since the background solution is periodic in time  we have to resort to Floquet theory to investigate the linear stability. We choose
\begin{eqnarray}
\tilde{b}(t,z) &=& e^{i\lambda t} \left( \sum_{n= 0}^{\infty} \cos(n\omega t) \tilde{b}^{(1)}_n(z) + \sum_{n= 1}^{\infty} \sin(n\omega t) \tilde{b}^{(2)}_n(z) \right),\label{eq_Floquet_TPS_b}
\\
\tilde{P}(t,z) &=& e^{i\lambda t} \left( \sum_{n= 1}^{\infty} \sin(n\omega t) \tilde{P}^{(1)}_n(z) + \sum_{n= 0}^{\infty} \cos(n\omega t) \tilde{P}^{(2)}_n(z) \right), \label{eq_Floquet_TPS_P}
\end{eqnarray}
as an ansatz, where $\omega$ is the frequency of the background TPS, and $\tilde{b}^{(1)}, \tilde{P}^{(1)}, \tilde{b}^{(2)}, \tilde{P}^{(2)}$ and $\lambda$ are unknown. The spectrum $\{\lambda_n(\omega), n \in \mathbb N\}$ of perturbations is obtained by solving the linear system subjected to the mentioned Dirichlet boundary conditions. This is performed  through the same pseudospectral method used to find the TPSs themselves (see appendix D in \cite{biasi2018floquet} for details). Solutions always come in pairs $(\pm \lambda_n, \tilde{b}^{(1)}, \tilde{P}^{(1)}, \pm \tilde{b}^{(2)},\pm \tilde{P}^{(2)})$. This entails that, as soon as  $\textrm{Im }\lambda_n \neq 0$ for some $n$, there is a linearized fluctuation that grows exponentially with time, i.e., a linear instability. On the contrary, if $\lambda_n \in \mathbb R$ for all $n$, the perturbations remain bounded and the TPS is linearly stable.

After a brief introduction to the required techniques, we are ready to present the main results we have obtained.
The TPSs we have constructed conform a surface in the three-dimensional space spanned by the driving frequency $\omega$ and the amplitudes of the gravitational squeezing at the asymptotic boundary, $\rho =  {\rm max}_t   |b(t,0)|$, and tip, $\rho_ t  = {\rm max}_t | b(t,z_0)|$. 

\begin{figure}[h!]
\begin{center}
\includegraphics[scale=0.4]{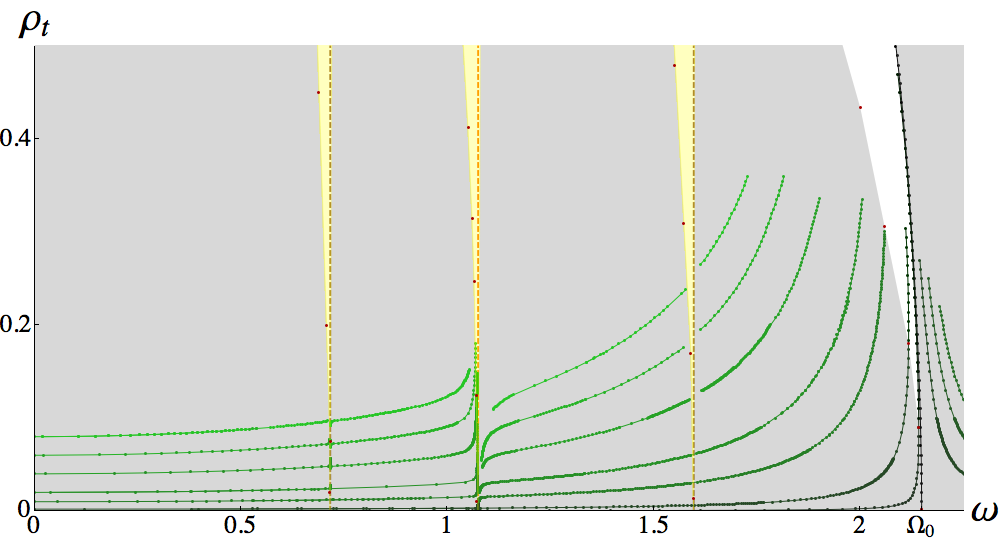}
\caption{\small  Points in the shaded region correspond to linearly stable TPSs. Solutions in the white region are linearly unstable; once perturbed, they evolve either to multiperiodic (even chaotic) geometries or black holes. Red points signal the data constructed through quasistatic quenches needed to determine the thresholds of shaded areas. Solid lines are level curves of constant  $\rho$. Emerging from  $\Omega_0$, we see the first line of
nonlinear gravitational normal modes (solid black), which corresponds to TPSs with  $\rho = 0$. Finally, dashed lines indicate zones where the frequency of the source, $\omega$, resonates with some of the frequencies of the normal modes, $\Omega_n$, the orange one $\omega = \Omega_0/2$ and the brown ones from left to right $\omega = \Omega_0/3, \Omega_1/3$. Around these resonances (yellow regions), no TPSs were found. More resonances are expected at $\omega = \Omega_n/k$, for $n,k\in \mathbb{N}$, nevertheless, they are so narrow that are difficult to determine.}
\label{fig:fullplot0}
\end{center}
\end{figure}

In Fig.\ref{fig:fullplot0} we summarize our results as a  plot in the $(\omega,\rho_t)$ plane. This figure shows the level curves $\omega(\rho_t)$ of constant $\rho$ (in green), the regions of linear stability (shaded) and instability (white), as well as wedges where TPSs were not found (yellow). Solutions for moderate values of $\rho_t$ ($> 0.25$) are very difficult to obtain. While solutions below $\rho_t < 0.25$ can be obtained reliably with pseudospectral methods, above this value, construction via a quasistatically modulated driving is needed (see next section). 

On the right-hand side of Fig.\ref{fig:fullplot0}, the black line signals the nonlinear gravitational normal (sourceless) modes. Similarly to the case of a scalar field, this line branches off from a linearized normal mode frequency, $\Omega_0$,  and tilts toward lower values of $\omega$. It also corresponds to the right boundary of an instability region (the white wedge). The solutions on the line are themselves linearly stable, and their energy density increases monotonically as we move upward. It is expected that this linear stability is lost when the energy density reaches an extremum, a point that would correspond to a Chandrasekar-like limit.\footnote{This is in line with previous observations for nonlinear normal modes of massless scalar fields in global AdS \cite{Maliborski:2016zlh}, boson stars \cite{biasi2018floquet}, and solitonic solutions in Einstein-Maxwell-scalar theory \cite{Arias:2016aig}.} Unfortunately, this region has remained unreachable to our pseudospectral codes. 

\begin{figure}[h!]
\begin{center}
\includegraphics[scale=0.5]{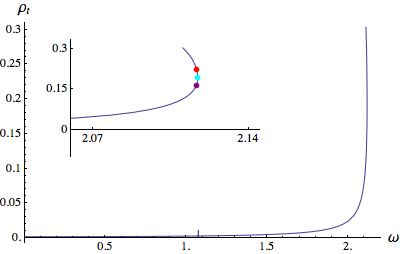}~
\includegraphics[scale=0.5]{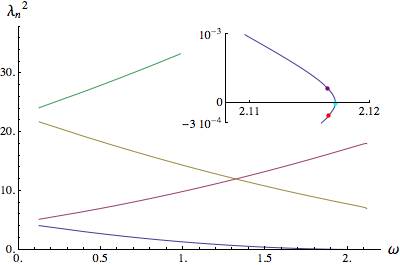}
\caption{\small  Left: Line of TPSs with $\rho = 0.001$. In the inset, purple, cyan and red dots mark solutions below, on and above the turning point respectively. Right: Floquet exponents $\lambda_0^2,\ldots,\lambda_3^2$ for the TPSs shown in the left plot. Solutions above the turning point become linearly unstable (Im$(\lambda_0)\neq 0$). In the inset, we zoom in this behavior. The color coding is the same as for the left inset.}
\label{fig:Linearstabilityplot}
\end{center}
\end{figure}

The other place where we have found a linear instability to set in is the turning point of the frequency in the level curves, $\omega(\rho_t)$, of constant $\rho$. In Fig. \ref{fig:fullplot0}  the locus of such points gives the left boundary of the instability region represented by the white wedge. 
In Fig.\ref{fig:Linearstabilityplot} we illustrate this phenomenon by tracking the lowest values of $\lambda_n$ for the curve of constant $\rho=0.001$.

The endpoint of the linear instability of the TPSs belonging to the white region of Fig.\ref{fig:fullplot0} can be explored numerically, by evolving the perturbed TPSs. When there is  a single exponentially growing Floquet mode ($\tilde{b}_0(t,z), \tilde{P}_0(t,z)$), we  prepare  the initial data at $t=0$ as follows
\beq
b(0,z) = b_p(0,z) + \epsilon \, \tilde{b}_0(0,z)\qquad, \qquad P(0,z) = P_p(0,z) + \epsilon \, \tilde{P}_0(0,z). \label{initial_data_perturbations}
\eeq 

\begin{figure}[h!]
\begin{center}
\includegraphics[scale=0.35]{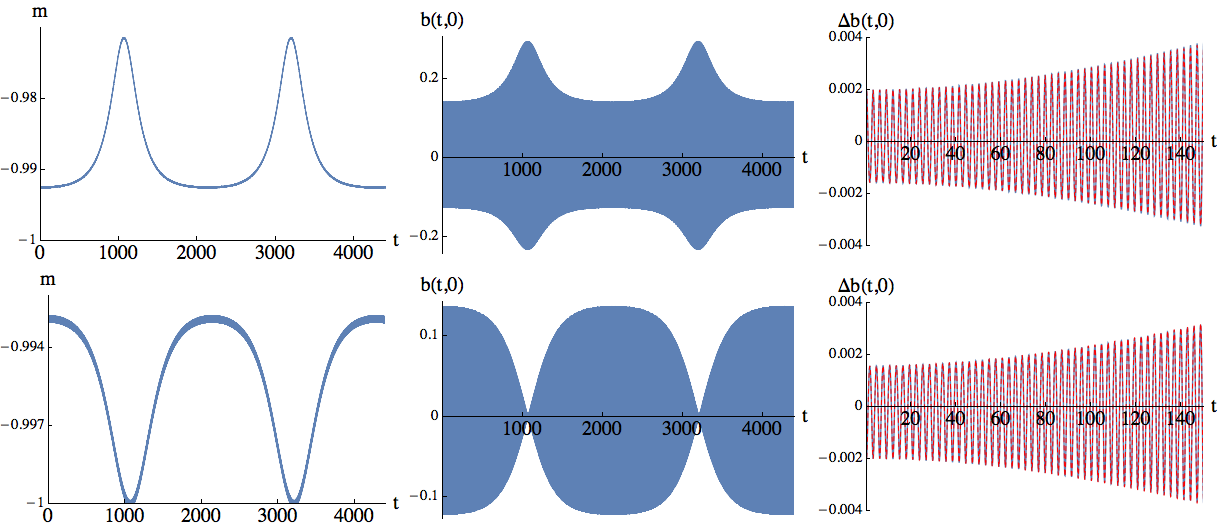}
\caption{\small  Numerical simulations of the initial data (\ref{initial_data_perturbations}) for the TPS: $\rho = 10^{-4}$, $\rho_t = 0.14$ and $\omega = 2.137$. The upper plots correspond to $\epsilon = 10^{-3}$ and the lower ones to $\epsilon = - 10^{-3}$. The plots on the left show the evolution of the energy density, the middle ones the evolution of $b(t,z_0)$. On the right, we compare the difference $\Delta b(t,z_0) = b(t,z_0) - b_p(t,z_0)$ (in blue) with the linearized unstable normal mode $\tilde{b}_0(t,z_0)$ (in dashed red), showing full agreeement.}
\label{fig:Perturbations_two_signs}
\end{center}
\end{figure}

If the unstable TPS is sufficiently close to the stability region, the evolution does not necessarily lead to gravitational collapse. In general, and as the example provided in Fig.\ref{fig:Perturbations_two_signs} illustrates, the final outcome depends on the sign of $\epsilon$. As we can see in this figure, the endpoint of the linear instability corresponds to a geometry that, in addition to the modulation associated with the external driving, develops additional long-time pulsations. We shall refer to these regular geometries as \emph{time-modulated soutions} (TMSs). In the case at hand, the sign of $\epsilon$ determines which particular kind of TMS the system relaxes to: for $\epsilon > 0$, the TMS pulsations increase the energy density while, for $\epsilon < 0$, they decrease it. The fact that in this example the system does not undergo gravitational collapse is probably related to the weakness of the external driving. For unstable TPSs associated with higher driving amplitudes, the $\epsilon > 0$ perturbation is expected to lead to black hole formation. 

%
%
As of now, our findings are in perfect agreement with the results obtained in \cite{biasi2018floquet}. However, there is one particular phenomenon which was overlooked there. For specific driving frequencies, the TPS resonates with higher normal modes, and this resonance triggers a new kind of behavior. In the limit of weak nonlinearities, $ \rho_t \ll 1$, a straightforward argument for the existence of these resonances is as follows. Recall that, from the perturbative analysis, all multiples of the driving frequency $\omega$ are excited. If this sequence hits some normal mode frequency, i.e, if $ k \omega = \Omega_n$, then the forcing term will contain a solution to the homogeneous equation, $\rho = 0$, leading, as usual, to the appearance of secular terms that may drive the system far away from the original TPS.\footnote{This argument can also be applied to a  massless real scalar field in global AdS$_4$. In this case, the normal mode frequencies are given by $\Omega_n = 3+2n$, and the Fourier modes of TPSs are restricted to odd multiples of $\omega$, so we expect the strongest resonance at $\omega = \Omega_0/3 = 1$, which is outside the region studied in \cite{biasi2018floquet} (see Fig.33 in this reference).} 
In Fig.\ref{fig:fullplot0} the leading resonances are plotted,  along with the narrow wedges emerging from them (in yellow); in these regions, we have been unable to find TPSs. For the part where the spectral code works properly, we have observed that the width of the wedges decreases with increasing $n$ and $k$, the widest one being at $\omega = \Omega_0/2$. 
%
%

\begin{figure}[h!]
\begin{center}
\includegraphics[scale=0.26]{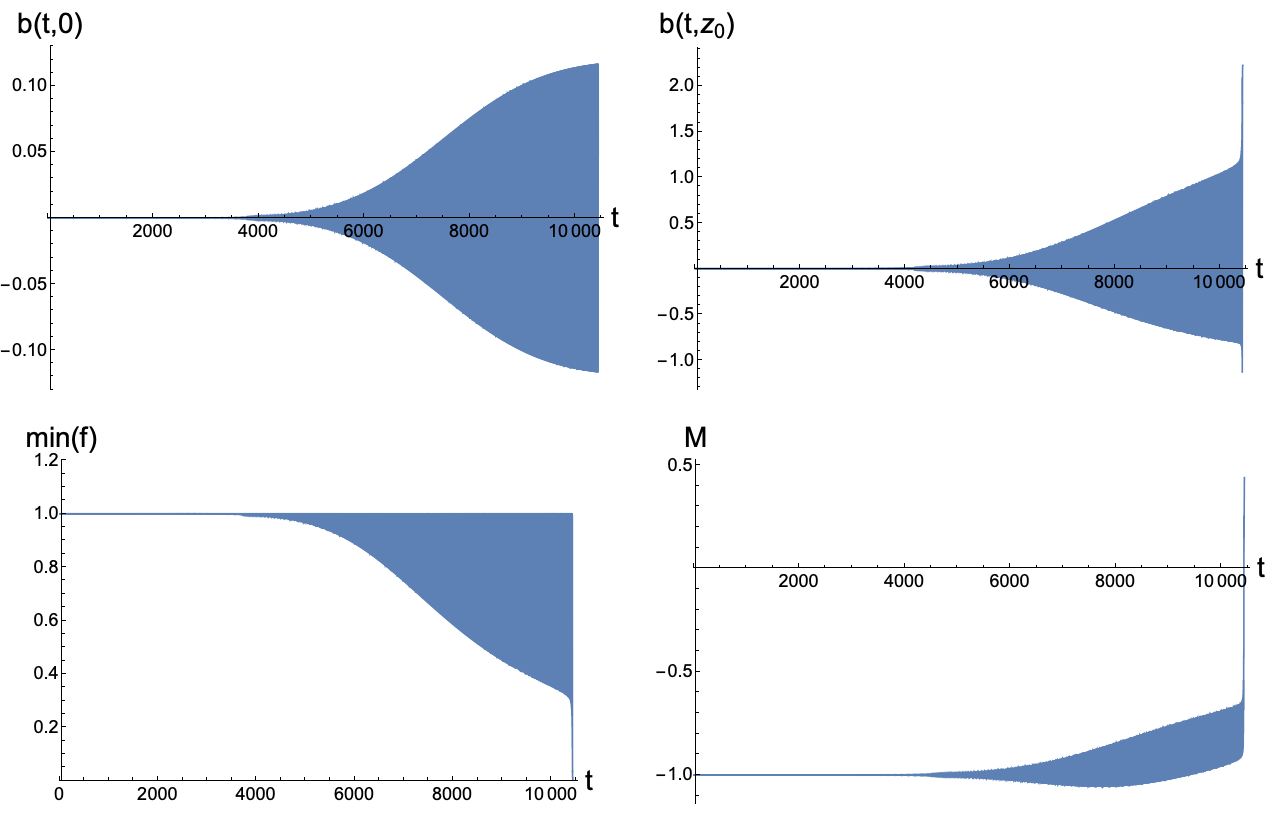}
\end{center}
\caption{\label{fig:collapseplots} \small Numerical simulation of a quasi-static build-up protocol with $\omega = 1.6$ and time span $\beta = 1.5\times  10^4$. In the upper two plots $b(t,0) = b_0(t)$ at the boundary and $b(t,z_0)$ at the tip are shown. The oscillations are so dense that they cannot be appreciated. All magnitudes evolve  adiabatically following a series of TPSs until a sharp threshold at $b_0(t_{col}) = 0.117$ is reached. There the instability sets in, as reflected in the sharp gain of net energy density $m$, and the formation of an apparent horizon, signalled by the function min$_z f(t,z)$ approaching zero. The momentum constraint is satisfied to a part in $10^{-4}$  throughout the whole process.}  
\end{figure}

\begin{figure}[h!]
\begin{center}
\includegraphics[scale=0.25]{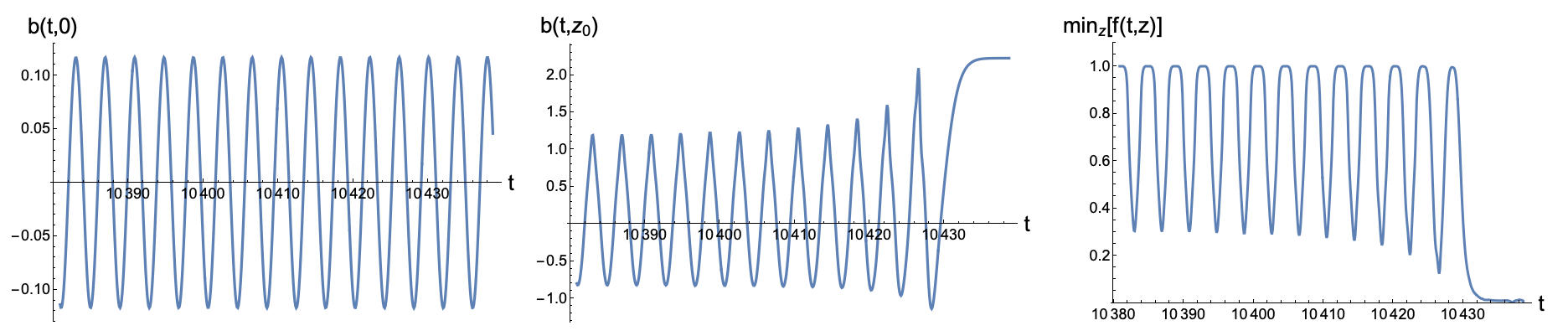}
\end{center}
\caption{\label{fig:collapseplotsenh} \small The same scenario depicted in Fig.\ref{fig:collapseplots}, where the final part of the evolution is shown. The extreme slowness of the amplitude increase contrasts with the abruptness of the sudden collapse that occurs in a few oscillations. The anharmonicity of the oscillation, as well as the lack of symmetry $b\to -b$, can be seen in the plot of $b(t,z_0)$ at the tip.} 
\end{figure}

\section{Modulated drivings}
\label{sec3}


In this section, the original motivation was to explore the transition to decoherence. In the meantime, we unraveled another transition that occurs at lower amplitudes.  Decoherence is dual to the formation of a black hole, whose horizon $z_h$ covers the tip of the cigar, $z_h<z_0$. Taking this fact into account, we will consider a periodic driving of frequency $\omega$, for which we will try to determine what is the maximum amplitude $\rho$ that the system can support without undergoing gravitational collapse. As discussed in section \ref{sec1}, if this transition takes place for $\rho \ll 1$, we may think of it as caused by the passage of a gravitational wave. 

The search, in this and the following subsection, is carried out numerically, by implementing a protocol that, starting from the static AdS-soliton, slowly amplifies a harmonic driving with a fixed boundary frequency $\omega$. The particular amplitude modulation we will employ is given by the following function interpolating between $\rho=0$ and $\rho = \rho_f$:\footnote{The non-analiticity at the origin, namely the fact that $\rho^{(n)}(0)=0 ~\forall n $, is essential to fulfill the boundary conditions at $t=0$ and, therefore, avoid introducing noise.}
\beqa
\rho(t) &=& \frac{\rho_f }{2}\left(1-\tanh\left(\frac{\beta}{t} + \frac{\beta}{t-\beta} \right) \right) ,~~~~~0\leq t<\beta, \nonumber \\
\rho(t)  &=& \rho_f  \, ~~~~~~~~~~~~~~~~~~~~~~~~~~~~~~~~~~~~~~~~~~~t \geq \beta 
\label{source_profile}
\eeqa
This corresponds to the upper envelope of the top left plot in Fig.\ref{fig:collapseplots}, where $b_0(t) = \rho(t) \cos \omega t$ is shown. The oscillations are so dense that they fill the shaded blue area.

The ramping parameter $\beta$ sets the time scale for reaching the final amplitude $\rho_f$. In our numerical searches, $\beta$ will be taken large as compared with the driving period, e.g., $\beta = {\cal O}(10^3/\omega)$. We will refer to these build-up processes of the final driving as {\em quasi-static}. For them, the system initially follows the increase in driving amplitude adiabatically, i.e., by going through a succession of TPSs. After reaching some amplitude, this adiabatic response might stop holding, and our task will be to map out when this occurs. In order to so, we will fix $\omega$ to a given value and take $\rho_f$ large enough so as to ensure that the system undergoes gravitational collapse at a 
given $\rho_{col} <\rho_f$. The reader can find an example of one such quasi-static build-up protocol in figures \ref{fig:collapseplots} and \ref{fig:collapseplotsenh}. 

\begin{figure}[h!]
\begin{center}
\includegraphics[scale=0.35]{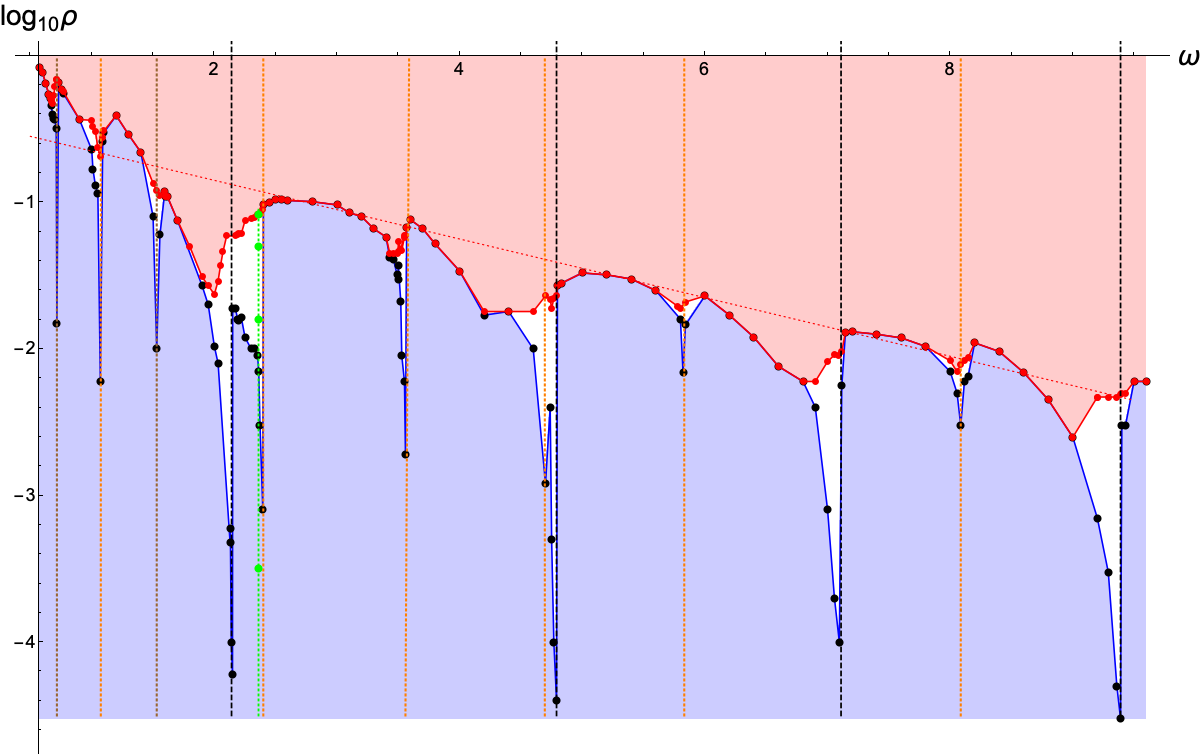}
\caption{\small   In blue, regions where TPSs are reached adiabatically from the AdS-soliton (at the bottom) by means of a slow build-up of the driving amplitude (vertical path).
 Traversing the blue curve the solution stops being a TPS. Points in the white region correspond either to multiperiodic or chaotic solutions. The vertical dashed lines indicate
 the frequencies of the linearized normal modes. Their halves, $\Omega_n/2$, are signaled by dotted vertical orange segments. In brown, $\Omega_0/3$ and $\Omega_1/3$ (see Fig.\ref{fig:fullplot0}). In green, the build-up protocol shown in  Fig.  \ref{fig:collapseplots2} with the snapshots of in Fig. \ref{fig:snapshots}. The red dotted diagonal is a fit to the highest values and decreases as $\rho \sim 10^{-\omega/5}$.}
\label{fig:stabilityplot}
\end{center}
\end{figure}

In Fig.\ref{fig:stabilityplot} we explore the threshold for collapse. The blue area contains the TPSs that are accessible from the AdS-soliton by the build-up protocol we have described. 
The blue line signals the loss of adiabaticity through a strong mixing with higher harmonics of the drive.  This leads to a non-periodic and even chaotic response. Remarkably,  the system does not undergo gravitational collapse and resists an increasingly intense driving until the red curve is crossed. The resonance wedges in this plot are in precise correspondence with the ones shown in yellow in Fig.\ref{fig:fullplot0}. 

\begin{figure}[h!]
\begin{center}
\includegraphics[scale=0.32]{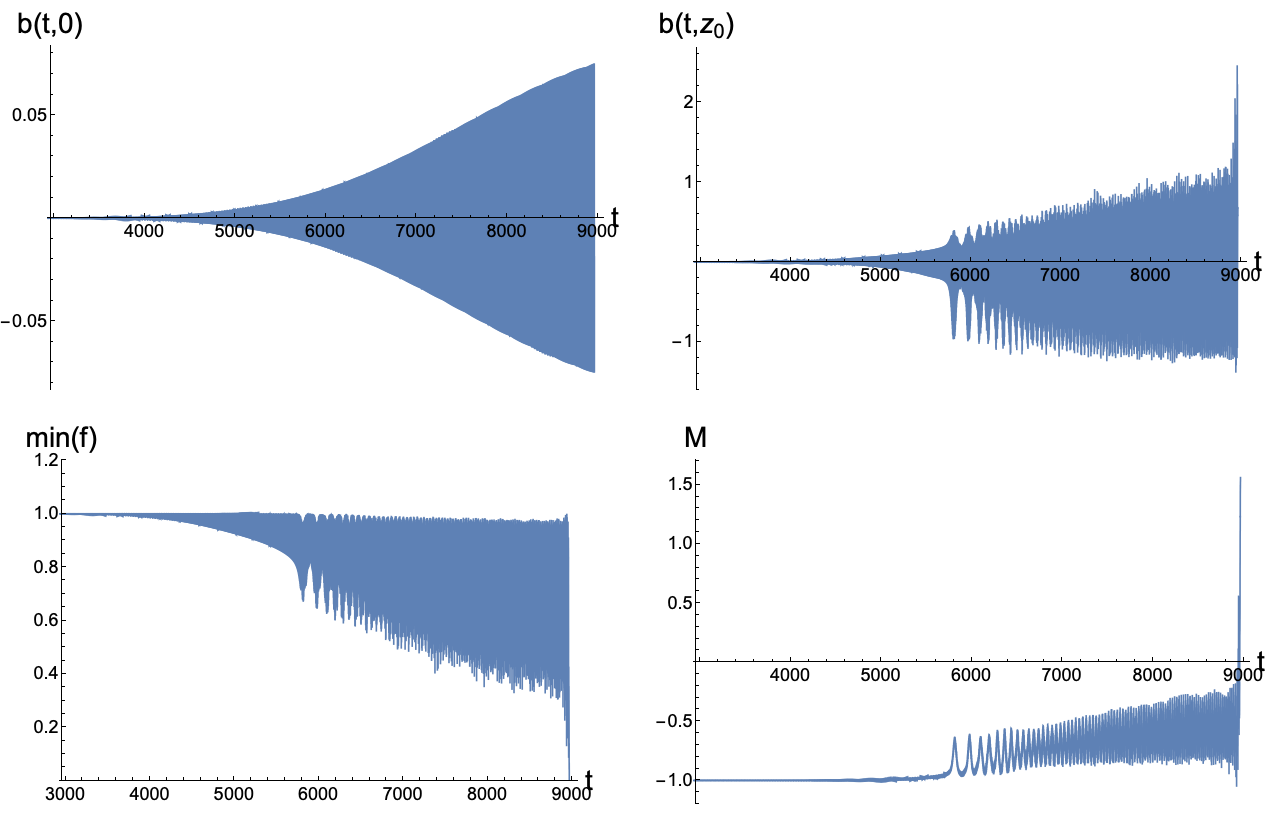}
\end{center}
\caption{\label{fig:collapseplots2} \small Numerical simulation of a quasi-static build-up protocol with $\omega = 2.375$ and a time span $\beta = 1.5\times  10^4$ (see green dotted line in Fig.\ref{fig:stabilityplot}).
All magnitudes evolve smoothly until an abrupt change is observed at $t \approx 5800\,$ with  $b_0 \approx 0.013$.  This transition signals the coupling to higher normal  modes.  The momentum constraint is always satisfied within one part in $10^{-4}$.}
\label{fig:nonadiab}
\end{figure}

\begin{figure}[h!]
\begin{center}
\includegraphics[scale=0.25]{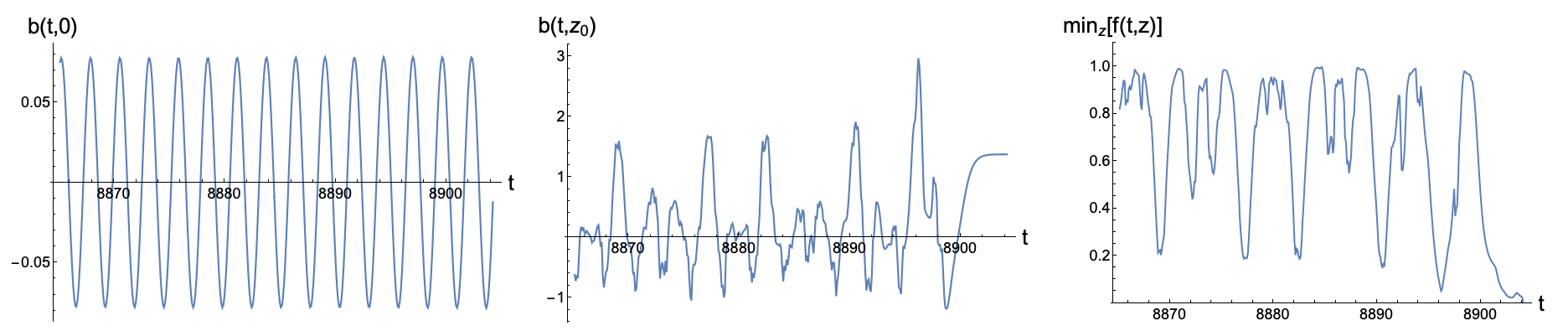}
\end{center}
\caption{\label{fig:chaoticplot} \small The same scenario depicted in Fig.\ref{fig:collapseplotsenh},  now  for the protocol  at  $\omega = 2.375$ of Fig. \ref{fig:nonadiab}. The evolution for $t >8000\, $ becomes chaotic, as shown by the time evolution of the $b(t,z_0)$ at the tip. }
\end{figure}

In Fig.\ref{fig:nonadiab} we can see the evolution of several magnitudes as the amplitude ramps up from zero with $\omega=2.375$ and crosses the blue line. When $\rho(t)$ reaches $0.013$, the geometry shows a strong modulation, as illustrated by the behavior of $b$ at the tip, $b(t,z_0)$, and the minimum of the emblackening factor, min$_zf(t,z)$. This is a clear signal of interference among two or more frequencies. In Fig. \ref{fig:chaoticplot} we plot the time evolution of two quantities associated with the bulk interior,  $b(t,z_0)$ and min$_z[f(t,z)]$, in the last time lapse before collapse. 
As compared with the TPS case in Fig. \ref{fig:collapseplotsenh}, we see the interior evolution becoming chaotic. This, however, has been sustained without horizon formation since $t\sim  8000$  at least,\footnote{ By $t$ we will hereafter implicitely imply the pure number $t/z_0$.} corresponding to a sizeable window in driving amplitude. 

\begin{figure}[h!]
\begin{center}
\includegraphics[scale=0.25]{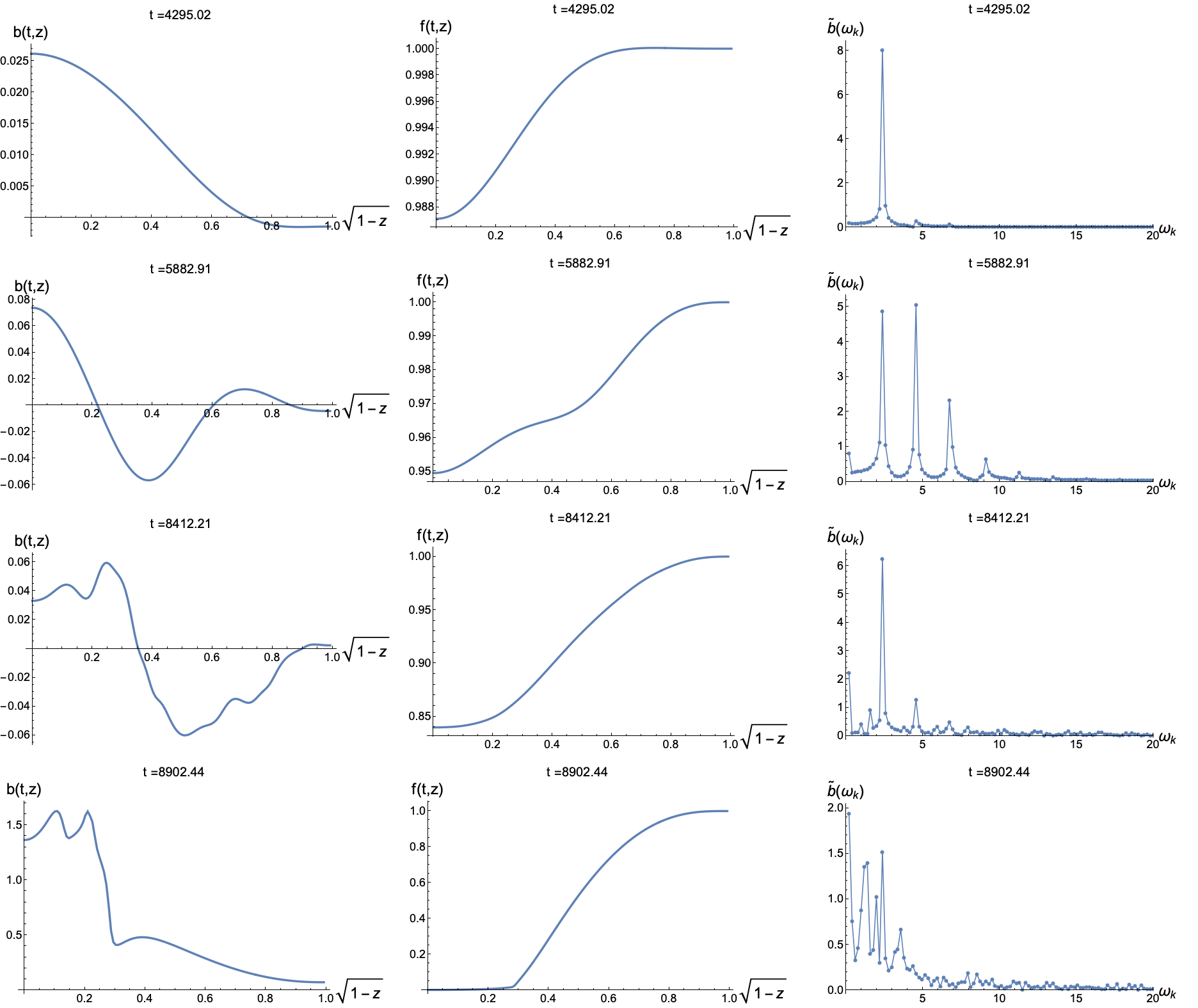} 
\end{center}
\caption{\label{fig:waveforms} \small Four snapshost of the build-up protocol of Fig.\ref{fig:collapseplots2} corresponding to  $\omega = 2.375$ (see green dots  in Fig.\ref{fig:stabilityplot}). The abscisa has been taken to be $\sqrt{z_0-z}$ with $z_0=1$ in order for the profiles to range from the tip (on the left, $z=1$) to the asymptotic boundary (on the right, $z=0$). The Fourier transform of $b(t,z_0)$ at the tip, over a time span of $\Delta t = 400$,  is shown on the right.  The resonant coupling to the higher normal  modes is apparent at $t=8412\,$. For higher driving amplitudes a cascade towards higher frequencies is behind a wiggly, yet regular, profile for $b(t,z)$.  Finally, the apparent  horizon starts forming near $z=0.3$ as seen in the last plot of $f(t ,z)$.}
\label{fig:snapshots}
\end{figure}

In order to make this statement more precise, we show four snapshots of the fields $b$ and $f$  in Fig.\ref{fig:snapshots}; to the right,   we represent the Fourier transform of a sequence containing 400\, seconds of the driving. In the first row, at $t=4295\,$, i.e. for low amplitude, the TPS Fourier content is solely given by the driving frequency. At $t=5882\,$ a clear resonant enhancement of the higher normal modes is observed. 
Indeed, normal modes $\Omega_2 = 4.790 \sim 2\omega$ and $\Omega_3 = 7.11 \sim 3\omega$ are natural higher harmonics of the driving frequency.\footnote{This situation bears a strong resemblance with the multioscillator solution in \cite{Choptuik:2018ptp}. Here this  solution lives {\em on top} of a sourced TPS, which triggers it  resonantly at a concrete frequency and amplitude window.} In the third row, at  $t=8412\,$,  the dynamics looks again dominated by the (now more intense) driving. However it starts populating all frequencies, the shape becomes irregular, and the motion chaotic, as seen in Fig. \ref{fig:chaoticplot}. Finally, at $t=8902\,$, an apparent horizon starts forming. Past this point, and since the external driving remains active, the system is expected to keep absorbing energy while the apparent horizon grows without bound. Following \cite{rangamani2015driven}, a set of different regimes will take over as for the inexorable temperature growth of the quantum system.

Before moving on to the next subsection, let us discuss briefly the numerical methods we have employed to perform the simulations. The radial direction has been discretized by a finite-difference scheme that employs fourth-order accurate centered stencils; typically, our grid contains $2^{10}+1$ points on the interval $0 \leq z \leq 1$. Likewise, the time evolution has been performed with an explicit fourth-order Runge-Kutta method. 
At the origin, the only regularity condition imposed is $f(t,z_0) = \exp(- b(t,z_0))$, which avoids a conical singularity. For the radial coordinate, following \cite{Craps:2015upq}\cite{Myers:2017sxr}, a convenient change of variables $r = \sqrt{z_0-z}$ has been employed to treat
correctly the vicinity of the tip $r=0$. Despite the rather low resolution used, the need for high computational resources came from the long duration of the build-up protocols studied. Therefore, we have employed two different parallel codes to allow cross-check: one in Fortran 90 (parallelized with MPI) and another one in C (parallelized in CUDA, and run on the Nvidia TeslaK80 GPU). Both codes have been run at CESGA.\footnote{Centro de Supercomputaci\'on de Galicia, \url{http://www.cesga.es/}.}

\subsection{The non-quasi-static case}

In the previous subsection, we have focused on build-up protocols with $\beta \gg 1/\omega$, for which the driving amplitude increases very slowly toward its final value. This was motivated by the observation that, for such quasi-static processes, the system responds adiabatically, passing through a succession of time-periodic solutions with the correct instantaneous boundary conditions. Conversely, if we depart from this limit, the response of the system stops being adiabatic, and other time-dependent geometries are excited.

In our previous work \cite{biasi2018floquet}, the same situation was explored for massless scalar fields in global AdS$_4$. It was found that, for driving frequencies $\omega$ sufficiently close to the linear instability line, a remarkable phenomenon took place. Specifically, there existed a critical build-up time span, $\beta_c$, at which the system underwent a sharp transition between two radically different late-time regimes. On the one hand, for $\beta > \beta_c$, a regular solution with a sharply pulsated periodic modulation was reached, which was termed {\em time modulated solution} (TMS).  Being horizonless, TMSs are dual to quantum coherent excited states. On the other hand, for $\beta < \beta_c$, gravitational collapse took place. The most surprising aspect of this transition is that as $\beta \to \beta_c^\pm$ (and at times $t > \beta$), the system spent a progressively longer time around an intermediate attractor, which turned out to be nothing but the linearly unstable TPS associated with the final driving. This observation allowed us to identify this transition as a type I critical phenomenon in gravitational collapse \cite{Gundlach:2007gc}. In this context, the novelty of this transition stems from the fact that it is not triggered by varying smoothly a one-parameter family of initial data \cite{Choptuik:1996yg}\cite{Bizon:1998qd}; instead, it appears upon varying a one-parameter family of time-dependent boundary conditions for the scalar field. This fact justifies employing the adjective {\it dynamical} to describe it. 

The purpose of this subsection is to demonstrate that type I critical phenomena also show up in the present setup. We will discuss two transitions in detail: one between two different kinds of TMSs, another between a TMS and a collapsing geometry.\footnote{The simulations to be discussed next have been performed on a grid with $2^{11} + 1$ points.} The averaged mass per period, 
\beq
\left< m \right>_T(t) \equiv \frac{1}{T} \int_{t-\frac{T}{2}}^{t+\frac{T}{2}}dt' m(t'),  
\eeq
will be a central quantity in our analysis. 

\begin{figure}[h]
\begin{center}
\includegraphics[width=15cm]{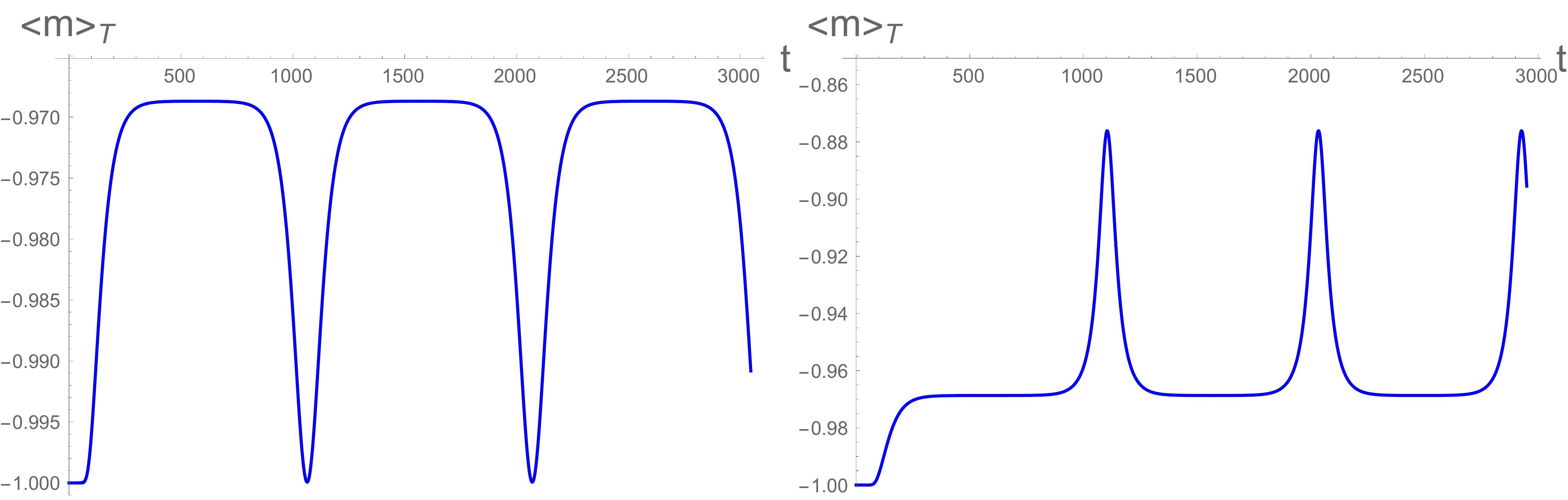}
\end{center}
\caption{\label{fig:up-down_2} {\small Left: For $\beta = 142.4117050 > \beta_c$, we obtain a TMS that beats downward. Right: for $\beta = 142.4116974 < \beta_c$, we obtain a TMS that beats upward.}}
\end{figure}

\begin{figure}[h!]
\begin{center}
\includegraphics[width=11cm]{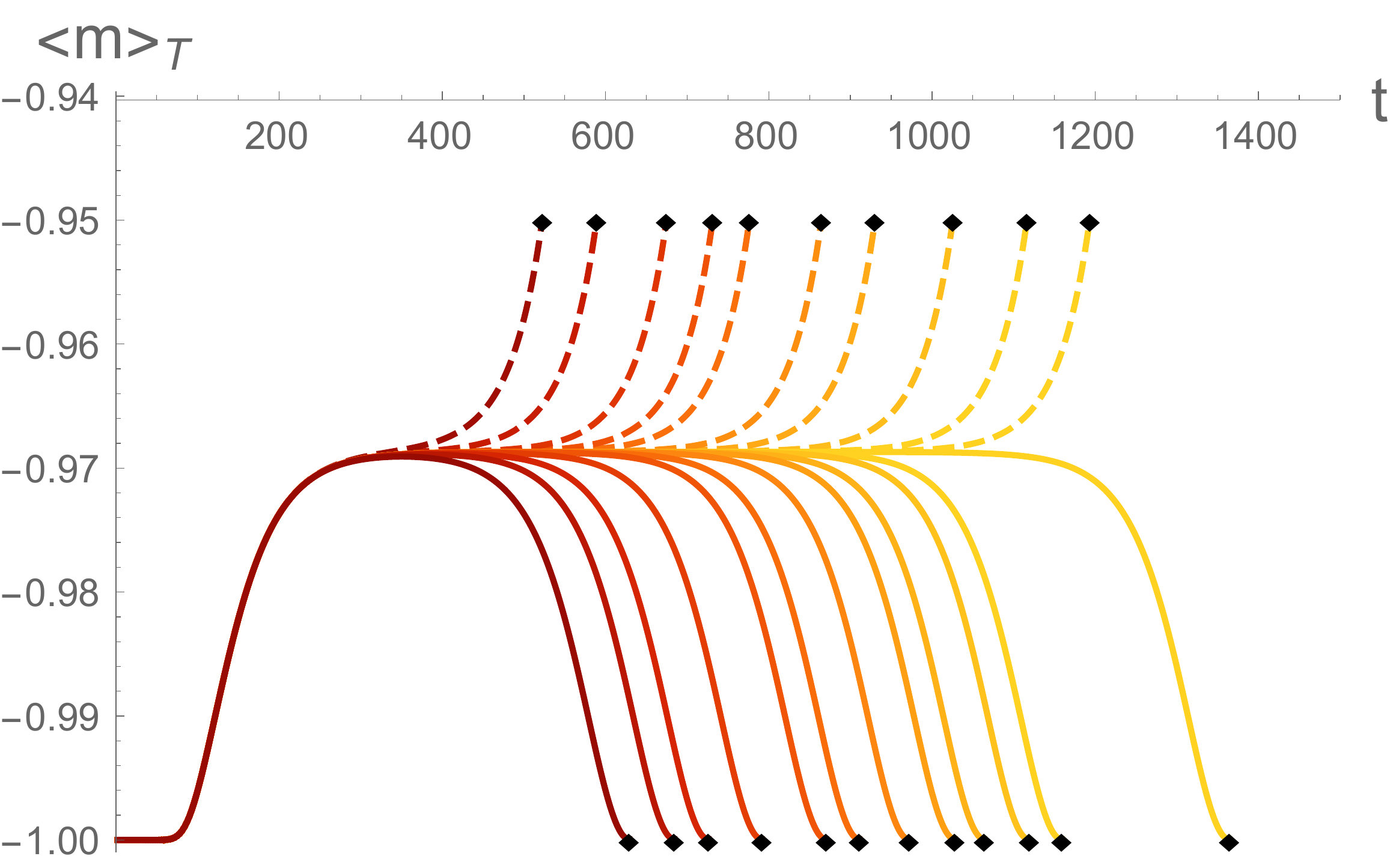}
\end{center}
\caption{\label{fig:strech_2} {\small Build-up processes leading to a TMS that starts pulsating upward (dashed) and downward (solid) as $\beta \to \beta_c$. Dashed simulations have been stopped when $\left< m \right>_ T = - 0.95$ for the first time. On the other hand, solid simulations have been cut when $\left<m\right>_T$ reaches its first minimum after the build-up phase.}}
\end{figure}

For the first case, we consider a family of build-up protocols that interpolate between the AdS-soliton vacuum at $t = 0$ and a driving  amplitude $\rho_f = 0.001$ at $t = \beta$ at fixed frequency  $\omega = 2.11$. As Fig. \ref{fig:up-down_2} (left) shows, for $\beta > \beta_c$ the system flows to a TMS  whose trademark property is having an averaged energy density per period $\langle m \rangle_T$ that, as $\beta \to \beta_c^+$, develops extended plateaux separated by fast downward beats. On the other hand, as illustrated in Fig.\ref{fig:up-down_2} (right), for $\beta \to \beta_c^-$ the TMS is characterized by a $\langle m \rangle_T$ that has the same plateau value, but starts beating upward.\footnote{It turns out that, as $\beta \to \beta_c^-$, the system goes from TMSs that beat alternative upward and downward to TMSs that only beat upward. Fig.\ref{fig:up-down_2} shows an example only of the latter situation.}
By looking at Fig.\ref{fig:strech_2}, it becomes obvious that, as $\beta \to \beta_c^\pm$,  the duration of the initial plateau after the build-up phase gets progressively longer. 

If the transition between the two kinds of TMSs we have found is to be interpreted as a type I critical phenomenon, the following relation must hold \cite{biasi2018floquet}
\begin{equation}
\Delta t = - \frac{1}{\lambda} \log |\beta - \beta_c| + a,  \label{master_relation}
\end{equation}
where $\Delta t$ is the permanence time around the unstable TPS, $\lambda$ is the norm of the purely imaginary eigenfrequency of this TPS, and $a \in \mathbb R$. The permanence time $\Delta t$ is best thought of as the length of the plateau; however, note that a shift in $\Delta t$ can be compensated by a change in $a$, leaving the values of $\beta_c$ and $\lambda$ invariant. This implies that, strictly speaking, there is no preferred definition of $\Delta t$: any two choices related by a shift are equivalent at the level of extracting $\beta_c$ and $\lambda$. Taking advantage of this freedom, for simplicity we define $\Delta t$ as follows.\footnote{We have checked that other possible definitions lead to compatible results.} First, for build-up processes that end in a TMS that starts beating upward, we define $\Delta t$ as the smallest time at which $\left<m\right>_T$ equals a particular predefined value $m_0$ above the plateau. In this example, we have chosen $m_0 = -0.95$. On the other hand, for build-up processes that end in a TMS that beats downward, we define $\Delta t$ as the time at which $\left<m\right>_T$ reaches its first minimum after the build-up phase. 

Applying the relationship \eqref{master_relation} to simulations with $\beta > \beta_c$, we get that\footnote{A word of caution is in order. While it can be argued that the value of $\lambda$ obtained by this procedure is resolution-independent, this is not the case for $\beta_c$. A precise determination of $\beta_c$ requires us to take the double scaling limit $\beta - \beta_c \to 0$, $r \to \infty$, where $r$ is the number of discretization points.}
\begin{equation} 
\beta_c = 142.4116999 \pm 1 \times 10^{-7},~~~~~~\lambda = 0.0225 \pm 0.0002. 
\end{equation}
 For $\beta < \beta_c$, we obtain 
\begin{equation} 
\beta_c =142.4117002 \pm 3 \times 10^{-7} ,~~~~~~\lambda = 0.0220 \pm 0.0002. 
\end{equation}
Both values of $\beta_c$ and $\lambda$ are perfectly compatible with each other.

\begin{figure}[h]
\begin{center}
\includegraphics[width=15cm]{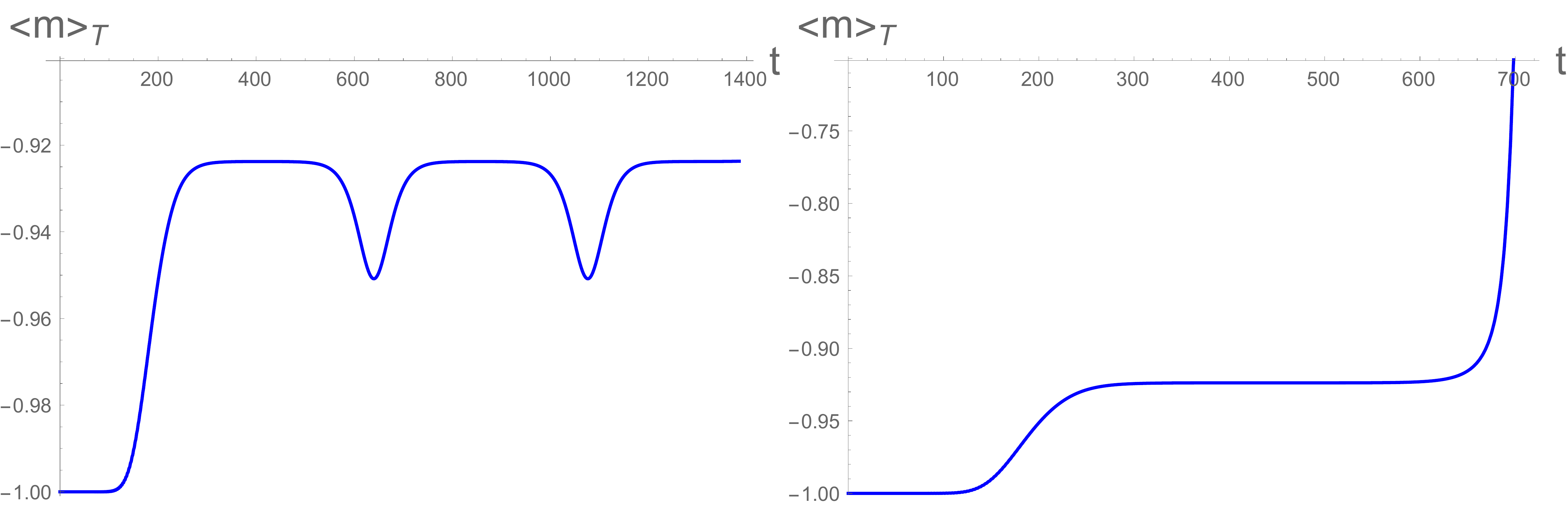}
\end{center}
\caption{\label{fig:up-down} {\small Left: For $\beta = 296.1971924 > \beta_c$, we obtain a TMS. Right: for $\beta = 296.1971680 < \beta_c$, the system undergoes gravitational collapse.}}
\end{figure}

\begin{figure}[h]
\begin{center}
\includegraphics[width=11cm]{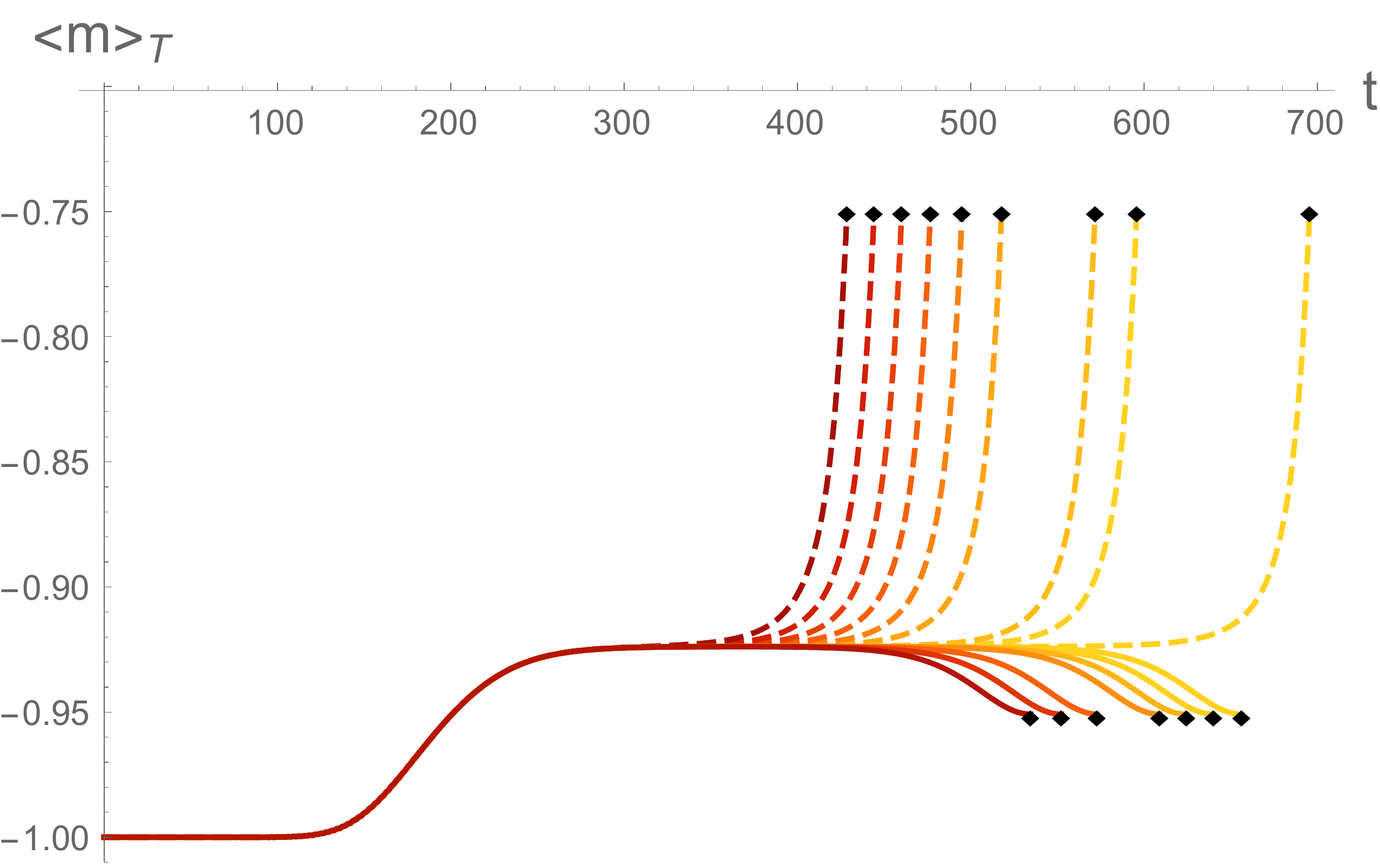}
\end{center}
\caption{\label{fig:strech} {\small Build-up processes leading to gravitational collapse (dashed) and a TMS (solid). Dashed simulations are only plotted until $\left<m\right>_T$ first reaches $-0.75$; solid ones until $\left<m \right>_T$ reaches its first minimum after the build-up phase.}}
\end{figure}

For our second example, we  set the driving frequency to $\omega = 2$ and the final amplitude to $\rho_f = 0.0114$. In this case, the system transitions from a collapsing geometry to a TMS as $\beta$ crosses $\beta_c$ from below. Relevant examples of such solutions are plotted in Fig. \ref{fig:up-down}. Again, for $\beta \to \beta_c^\pm$, the length of the initial plateau gets progressively longer (see Fig. \ref{fig:strech}). 

Repeating the fitting procedure described before,\footnote{For the collapsing branch, we define $\Delta t$ as the time for which $\left< m \right>_T$ first reaches $-0.75$.} we get that, for simulations with $\beta > \beta_c$, 
\begin{equation} 
\beta_c = 296.1971707 \pm 9 \times 10^{-7},~~~~~~\lambda = 0.0460 \pm 0.0004. 
\end{equation}
On the other hand,  for $\beta < \beta_c$, we obtain
\begin{equation} 
\beta_c = 296.197173 \pm 2 \times 10^{-6},~~~~~~\lambda = 0.04535 \pm 0.00005. 
\end{equation}
Again, both values of $\beta_c$ and $\lambda$ are compatible with each other. 

As a final consistency check, we would need to establish that the values of $\lambda$ extracted from these numerical simulations match the ones coming out of the pseudospectral code that computes the eigenfrequency spectrum of the unstable TPS. Unfortunately, both TPSs have remained outside the reach of our numerical construction methods. In this way, the values of $\lambda$ that we have quoted above have to be viewed as independent predictions that a yet-to-be-performed pseudospectral analysis would need to reproduce.  

\section{Conclusions and outlook}
\label{sec4}

In this work, we have pushed a step further the study of periodically driven quantum systems in the context of Holography. We have considered holographic CFTs with toroidal boundary conditions, where the periodic driving has been implemented by turning on a particular time-dependent shear deformation of the background metric in which these CFTs live. 

Our findings show that, for low amplitudes, the system allows for time-periodic solutions. We have uncovered the phase space of these geometries and discussed their stability, both at the linear and the nonlinear level. Globally, there is an overall limit to the amplitude of such periodic solutions, beyond which the system loses coherence, as signaled by the prompt collapse to a black hole. This threshold amplitude is an irregular function of the frequency decreasing on average as $10^{-\omega/5}$.

As compared with the case of a scalar field in global AdS$_4$ presented in \cite{biasi2018floquet}, the threshold amplitude here is lower, hence TPSs are now more fragile, something probably tied to the lack of reflection symmetry $b \to -b$ in the present case.
Apart from this, there are spiky wedges of instability at several discrete frequencies. For instance, close to the normal modes of the AdS-soliton, these instabilities can be seen to appear at the linearized fluctuation level. At frequencies that exactly divide the normal mode frequencies (dotted lines in  Fig.\ref{fig:fullplot0} and \ref{fig:stabilityplot}) they are intrinsically of nonlinear nature, originating from the resonant coupling to these modes. Inside the white wedges in Fig.\ref{fig:stabilityplot} this leads to modulated solutions and, for even higher amplitudes,  multi-oscillator bound states (see the second line in Fig.\ref{fig:waveforms}) and even chaotic evolutions. 
 
We have also studied the response of the system to modulated driving protocols outside the adiabatic regime, finding that, away from the quasi-static limit, the loss of adiabaticity can result in the same kind of dynamical type I gravitational phase transition found in other setups. 

Overall, the similarities between the results we have found and those presented in \cite{biasi2018floquet} suggest that periodically driven, finite-sized holographic systems feature several universal behaviors.

In the early days of the AdS/CFT correspondence, the AdS-soliton posed a puzzle due to its energy density being lower than that of AdS itself. This fact, instead of signaling an instability, was interpreted later as a Casimir energy, adding support to the picture of this geometry as being dual to a QFT on a compact domain (a torus). This raises the challenge of interpreting our results in the light of a Dynamical Casimir Effect \cite{dodonov2010current}. At particular values of the driving frequency and amplitude, the response becomes populated with the normal modes of the cavity. These modes are the poles that appear in the spectral function and would be the particles 
excited from the vacuum by the time-periodic boundary conditions.

Let us add some speculative remarks about the possibility of envisaging the studied process as the response of a two-dimensional strongly coupled quantum system of rectangular shape to the perpendicular passage of a gravitational wave. 
Resonant classical detectors are characterized by sharply peaked sensitivities around the normal mode frequencies. When cryogenized, at resonance they can detect strains  as low as $10^{-21}$ \cite{aguiar2011past}. In contrast, interferometers
have a much wider u-shaped sensitivity curve. The possibility of detecting resonantly via quantum effects  has been  discussed recently in different setups \cite{sabin2014phonon} \cite{Schutzhold:2018wfu}\cite{Robbins:2018thb}\cite{Landry:2016zip}.  Interpreting our results in a similar context, of quantum induced transitions, we would like to extract some gross features and generic lessons. The gravitational wave expected from binary coalescences has
a similar amplification in amplitude as the modulated drivings we have used in section \ref{sec3}.  At low amplitudes,  the response of the quantum system is an excited time-periodic coherent state (Floquet condensate). At some (frequency-dependent) threshold there is a sharp decoherence transition to a thermal state, as signaled by black hole formation in the bulk (see  Fig.\ref{fig:collapseplots}). In the absolute confidence that this transition has not been triggered by any other source  (perfectly isolated system),  we would term this transitions a  ``decohering detection".
At certain discrete frequencies, an intermediate transition to a regime of excited non-periodic coherent states populated by higher modes is seen (see Fig. \ref{fig:stabilityplot}). This takes place at much lower values of the amplitude, and we would call these ``resonant detections". 
 Led by this reasoning, we can assign to the plot in figure Fig. \ref{fig:stabilityplot} the meaning of a sensitivity plot, where the red and blue lines would correspond to the sensitivity curves for decohering and resonant detections respectively.
Our first generic conclusion would state that in a  gravitational wave detector based on a strongly coupled quantum system, the intrinsic nonlinearity introduces additional resonances at fractions of the natural normal mode frequencies, $\Omega_n/k$, which are thinner for higher $k$. We believe this is a model-independent statement.  
Another nontrivial outcome is that, at least for the range of frequencies examined,  there is an overall exponential increase in the sensitivity with frequency. Still within the range explored numerically, the minimum strains are too large, $\rho>10^{-4}$, compared with the relevant strains of interest.   It would require more computational effort to see whether this increase continues for higher frequencies or slows down. 
This observation is likely to be model dependent, hence calling for some holographic system that enjoys an enhanced sensitivity. A straightforward suggestion would be to explore the gravitational driving of a coherent Bose-Einstein condensate, like the one proposed in \cite{Nishioka:2009zj}, aimed at modeling an insulator-superconductor phase transition at zero temperature.


\section*{Acknowledgements}
We would like to thank Roberto Emparan,  Pablo Bueno, Ram\'on Massachs, and Alfonso V. Ramallo for discussions.

This work was supported by grants FPA2014-52218-P  and FPA2017-84436-P from Ministerio de Economia y Competitividad, by  Xunta de Galicia ED431C 2017/07, by FEDER and by Grant Mar\'\i a de Maeztu Unit of Excellence MDM-2016-0692. A.S. is happy to acknowledge support from the International Centre for Theoretical Sciences (ICTS-TIFR), Bangalore, and the Infosys International Collaboration Grant.  A.B. thanks the support of the Spanish program ''Ayudas para contratos predoctorales para la formaci\'on
de doctores 2015'' associated to FPA2014-52218-P. This research has benefited from the use of computational resources/services provided by the Galician Supercomputing Centre (CESGA).

\begin{appendix}

\section{Equations of motion}
\label{app_A}

In terms of the following variables\footnote{Here, primes correspond to $\partial_z$ and dots to $\partial_t$.} 
\beqa
B(t,z) = b'(t,z),~~~~~~P(t,z) =\frac{e^\delta}{f} \dot b\, ,
\label{eq_B_P_z}
\eeqa
the equations of motion can be casted in the form
\beqa
\dot B &=& \left(f e^{-\delta}  P \right)',  \label{eq_Bdot_z}\\
\dot P &=& \frac{z^2}{2} \left(  \left(  -3+2\frac{(1-z^3) }{ z^2 } B \right) f e^{-\delta} \right)',  \label{eq_Pdot_z}\\
f' &=& \frac{12}{z(4-z^3)} \left( 1 - f \right) +\delta' f,  \label{eq_f_z}\\
\rule{0mm}{7mm}
\delta' &=&\frac{z}{4-z^3}\left( -3z^2 B+(1-z^3)B^2 + P^2 \right).    
\label{eq_delta_z}
\eeqa
From \eqref{eq_Bdot_z}-\eqref{eq_delta_z} it can be seen that, at the linearized level, the squashing field $b(t,z)$ acts like a  massless scalar field. There is yet one additional equation given by the momentum constraint 
\be
\dot f = \frac{z}{z^3-4}\left( \rule{0mm}{4mm}(3z^2+2(z^3-1)B)P \right) f^2 e^{-\delta}, \label{momcons}
\ee
which is not independent from \eqref{eq_Bdot_z}-\eqref{eq_delta_z}. It's main purpose will be to provide a consistency check of our numerical results.   

Near the tip of the cigar, $z\sim z_0 = 1$, we have the following infrared series expansion
 \beqa
f(t,z) &=&  e^{-\tilde b_0(t)} + \tilde f_2(t) (1-z) + {\cal O}((1-z)^2), \label{expA} \nonumber\\
b(t,z) &=& \tilde b_0(t) + \tilde b_2(t) (1-z) + {\cal O}((1-z)^2),   \nonumber \\
\delta(t,z) &=& \tilde \delta_0(t) + \tilde \delta_2(t) (1-z) +  {\cal O}((1-z)^2),  
 \eeqa 
The boundary condition $f(t,1) = \exp(- b(t,1))$ has to be  imposed in order to avoid a conical singularity at the tip.

\end{appendix}


\end{document}